\documentclass{cjour}
\usepackage{epsfig}
\authorrunninghead{Willy Kley}
\titlerunninghead{Interaction between planets and disk}

\begin{document}

\title{The tidal interaction between planets and the protoplanetary disk}

\author{Willy Kley}
\affil{Theoretisch-Physikalisches-Institut, Universit\"at Jena,\\
Max-Wien-Platz 1, D-07743 Jena, Germany}
\email{wak@tpi.uni-jena.de}

\abstract{
The discovery of now about 20 extrasolar planets orbiting solar-type
stars with properties quite different from those in our Solar System
has raised many questions about the formation and evolution of planets.
The tidal interaction between the planet and the
surrounding disk determines the orbital properties and the mass of the
planet.

We have performed numerical computations of planets embedded in a
protoplanetary disk and found that for typical values of the
viscosity the planet may easily grow up-to ten Jupiter masses.
New results on the mass evolution and the migration of the planet
are presented.
}

\begin{article}

\section{Introduction}
\begin{figure}[ht]
\begin{center}
\epsfig{file=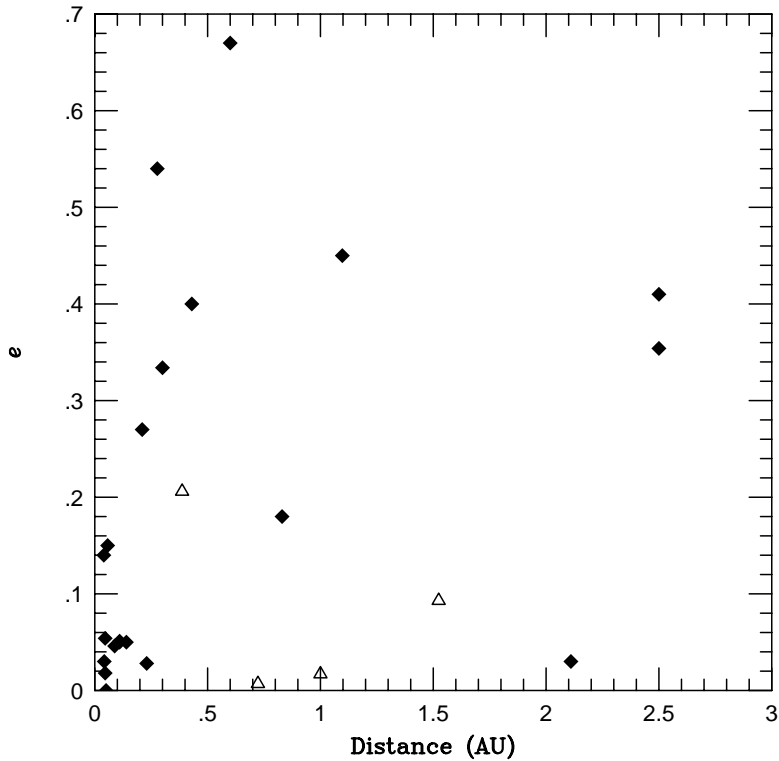, width=11cm}
\caption{\label{e-r.caption} Eccentricity versus semi-major
axis for the observed extrasolar planets around main-sequence stars (diamonds)
and the inner planets of the Solar System (open triangles).
Data are taken from \cite{encycl}.
}
\end{center}
\end{figure}

Since the discovery of the first extrasolar planet around a main
sequence star 5 five years ago \cite{mayor1995} the field of planet
searching has grown dramatically. Today about 20 twenty planets of this type
are known (for a summary see Marcy, Cochran \& Mayor 1999, \cite{marcy1999}).
The main difference to our own Solar System, where planets
have rather small masses and orbit the sun on nearly circular
orbits at distances of up-to several AU, has been the discovery of
massive planets (up-to 10 Jupiter masses, $M_{Jup}$). Some of the newly
discovered planets orbit their central stars very closely
(within a tenth of an AU) on orbits with smaller eccentricities. For larger
semi-major axis the eccentricities tend to be larger with a maximum of 0.67,
see Fig.~\ref{e-r.caption}.
The only extrasolar planetary system known
so far ($\upsilon$ And) consists of one planet at 0.059 AU on a
nearly circular orbit
and two planets at .83 and 2.5 AU having larger eccentricities (.18 and .41)
\cite{butler1999}.
For a general, up-to date review of the properties of the planets visit the
web-site of J. Schneider at {\tt http://www.obspm.fr/planets}.

These rather unexpected findings have brought new momentum into the
field of planet formation, and theorists have begun to explore the
physical conditions under which planets are supposed to form.
It is generally agreed that planets form simultaneously with the
star from a collapsing core of an interstellar gas cloud.
Angular momentum conservation leads to the formation of a disk
where mass is slowly transported inwards and accreted by the star.
Some of the material within this accretion disk condenses out
to form the planetary embryos which then grow during their final 
formation stages by rapid accretion of matter from the disk.
In this scenario all of the newly created planets orbit the stars
essentially on circular orbits as they originate from a Keplerian disk,
and the more massive, gaseous planets tend to have distances of several
AU from the star. 
While this scenario explains well the observational data of the Solar System,
it does not match so well the new results on the extrasolar planets.

As the last phase of planet formation determines the final masses
and orbital elements of the planets, it seems fruitful to concentrate in
particular on this evolutionary stage.
Here we present results of numerical computations of massive
(Jupiter type) planets embedded in and evolving simultaneously with
the protoplanetary accretion disk.

\section{Modeling planets in disks}
As the vertical thickness of accretion disks is very much smaller
than the radial extension, it is well justified to consider only
infinitesimally thin disks. Thus, working in cylindrical coordinates,
the problem is reduced to two dimensions ($r-\varphi$) where the disk
is located in the equatorial plane ($z=0$), and the origin
of the coordinate system is either centered in the middle of the disk or in the
center of mass of the system.
In this disk a one $M_{Jup}$ planet is placed initially
at a distance of $5.2$AU from the one solar mass star,
which is the present semi-major axis of Jupiter, $a_J$.

The initial surface density profile is given by $\Sigma \propto r^{-1/2}$
which is the analytic profile for a
constant viscosity disk with negligible radial infall velocity.
The total mass $M_d$ in the disk is
$10^{-2} M_\odot$. At the initial radial location of the planet $r=a_J$
an annular lowering of the density (gap) is imposed at $t=0$ \cite{kley1999}.
This speeds up the computations since through
the presence of the planet gravitational torques are exerted on the disk
which tend to open up an annular gap \cite{lin1993}.

Physically, the disk is treated as a viscous fluid with a standard
Reynolds stress tensor approximation. The form of the equations for this
problem is given for example in Kley (1999) \cite{kley1999}.
In these computations we use for simplicity a constant kinematic
viscosity which is $\nu = 10^{-5}$ in dimensionless units, where the
unit of distance given by the initial semi-major axis of the
planet $r_0=1 a_J = 5.2$AU,
and the unit of time is the inverse Keplerian period
$t_0= \Omega^{-1}_k(r_0)$ of Jupiter. In the graphics below time
is typically given in units the of corresponding orbital
period $P=2\pi t_0$ which is $11.9yr$ for Jupiter. 
The pressure is given by the assumption of a
constant vertical disk thickness $H/r = 0.05$ which identical to a constant
Mach number of 20.
We would like to point out that for the given vertical thickness the
viscosity corresponds to $\alpha=4 \cdot 10^{-3}$.
The basic model parameter are summarized in Table~\ref{modelpara.tab}.
\begin{table}[ht]
\caption{\label{modelpara.tab} The parameter of the standard disk model
in which the planets are embedded.
}
\begin{tabular*}{0.5 \textwidth}
{@{\extracolsep{\fill}}ll}
\hline
 Stellar Mass &   $1 M_{\odot}$  \cr
 Disk Mass    &   $0.01 M_{\odot}$  \cr
 H/r          &   $ 0.05$      \cr
 Viscosity &      $ 10^{-5} $      \cr
\hline
\end{tabular*}
\end{table}
 
Numerically, the combined system of a disk with an embedded planet is
modeled by a finite difference method \cite{kley1989} which is second
order in space and formally first order in time. Operator splitting is
used where the advection and force terms
are solved explicitly and the viscous terms implicitly.
Typically we use a grid size of $N_r \times N_\varphi = 128 \times 128$
gridcells which cover a ring-like area from $r=0.25-4.0$ corresponding
to $1.3-20.8 $AU.
Numerical experiments varying for example resolution and the extent of
the computational domain (and other numerical parameter) indicate
that the major results are not affected by numerical issues \cite{kley1999}.
Some of the computations where the planet remains on a fixed
circular orbit are performed in the co-rotating frame where the planet is
fixed in the grid. Other computations where the planet's orbital
parameter are allowed to vary are typically performed in the inertial
frame.
\section{Results}
\subsection{Fixed orbital parameter of the planet}
\begin{figure}[ht]
\begin{center}
\epsfig{file=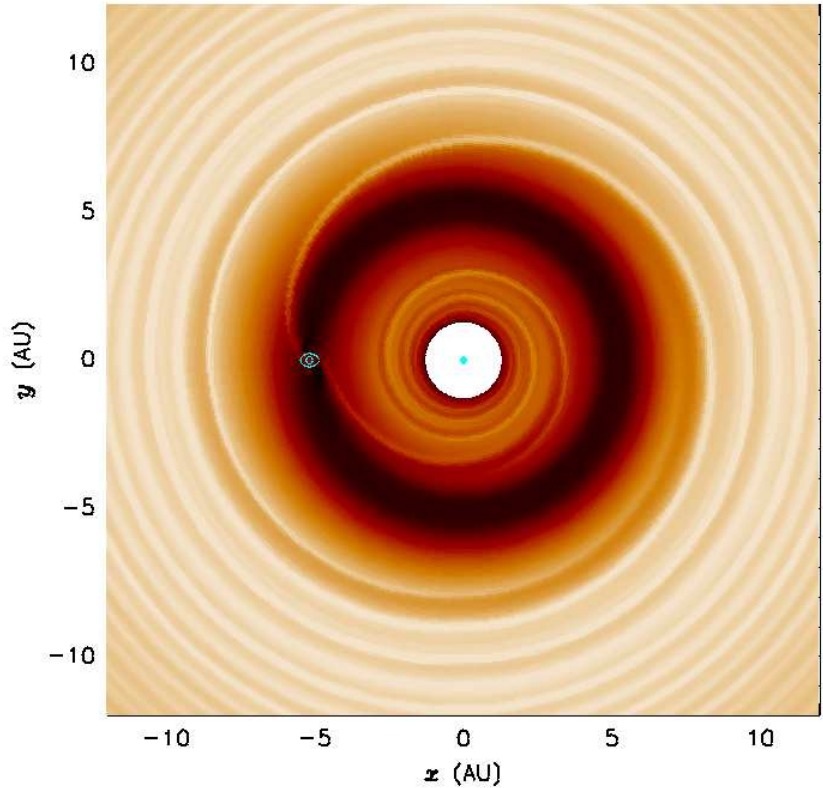, width=11cm}
\caption{\label{spirale.caption} Gray scale plot of the surface density
distribution after 200 orbits of the planet, which is located at
$x=-5.2, y=0.0$. The white line around the planet indicates its Roche-lobe.}
\end{center}
\end{figure}

\begin{figure}[ht]
\begin{center}
\epsfig{file=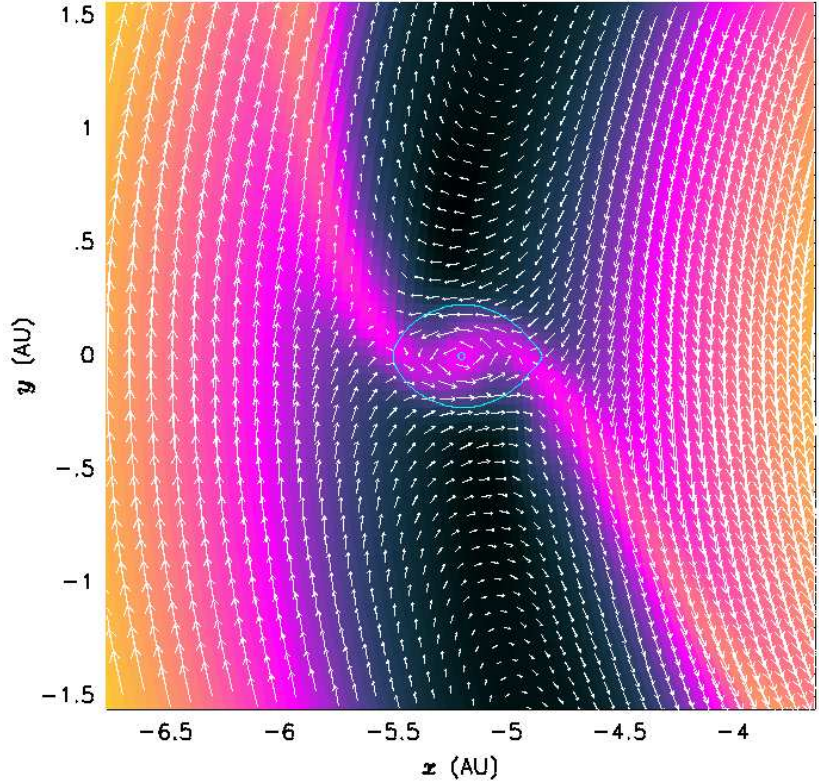, width=11cm}
\caption{\label{fig18au.caption} Gray scale plot of the surface density
distribution after 200 orbits of the planet, which is located at
$x=-5.2, y=0.0$.}
\end{center}
\end{figure}

First we consider the models where the planet was on a fixed circular orbit,
which means that the gravitational back-reaction of the disk onto the planet
was not taken into account. These models serve to extract the main
physical effects an embedded planet experiences within the disk.
An overview of the configuration of the standard model 
(see Table~\ref{modelpara.tab})
after about 200 orbits of the planet is presented in 
Fig.~\ref{spirale.caption},
where a gray-scale plot of the surface density is presented with lighter
color representing higher density. Clearly visible is the ring-like gap
created by the immersed planet where the density is about three orders
of magnitude smaller than in the surrounding disc. The gap is formed
because the planet acts to extract angular momentum from the inner
regions and transports if to the outer which effectively pushes
material away from the radial location of the planet.
Additionally, there are trailing
spiral shock waves induced in the disk. In the regions outside the planet
two spiral arms are inter-twined and inside there are three spirals,
where one always starts at the location of the planet.
The enlargement of the co-rotating flow-field for a higher resolution model
(with $448\times 128$ gridcells) in the vicinity of the
planet, given in Fig.~\ref{fig18au.caption},
shows the trailing character of the spiral arms. 
While the material close to $r=1$ encircles the star on horseshoe orbits
(in the co-rotating frame) some of it
enters the Roche-lobe of the planet from upstream region.
Mass from both edges, inner and outer, of the gap is able to enter
the Roche-lobe. 

Assuming that the matter which has entered the planet's Roche-lobe
becomes accreted by it, one can estimate the mass accretion rate onto the   
planet. For a total disk mass of $10^{-2} M_\odot$ 
this accretion rate is found to be
\[
\dot{M}_{pl} = 4.35 \cdot 10^{-5} M_{Jup}/yr
\]
which implies a mass doubling
timescale for a one Jupiter mass planet of only a few ten thousand years.
We note, that this accretion rate onto the planet is larger than
the equilibrium disk
accretion rate, $\dot{M}_{disk} = 3 \pi \nu \Sigma$, which implies that
for the given disk parameter all material which is moving viscously inwards
will eventually be accreted onto the planet.
This gas orbits the planet in the prograde direction in agreement with
the observed preferred direction of planetary rotation in the
Solar System.

\begin{figure}[ht]
\begin{center}
\epsfig{file=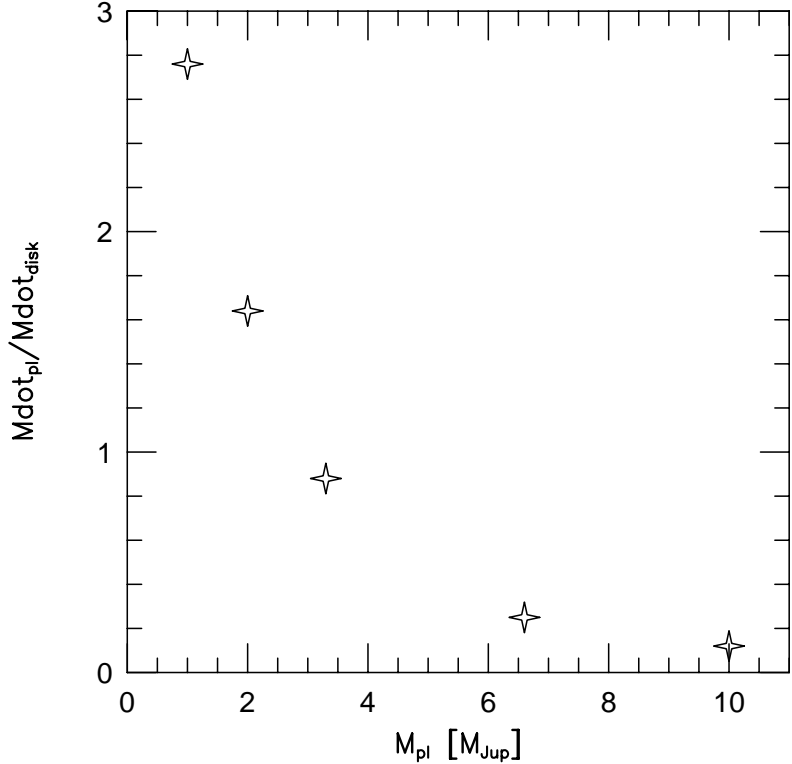, width=11cm}
\caption{\label{accq1.caption}
The normalized mass accretion rate $\dot{M}_{pl}/\dot{M}_{disk}$ onto
the planet versus the mass of the planet in units of $M_{Jup}$.
}
\end{center}
\end{figure}

However, the planet can not grow indefinitely since the accretion rate firstly
depends on the viscosity and, for values of $\alpha$ lower than about 
$4 \cdot 10^{-4}$ (a value probably not untypical for protostellar disks
which are only weakly ionized),
it becomes smaller than the disk accretion rate
\cite{kley1999, bryden1999}. Additionally, as $M_{pl}$ increases,
the torques exerted by the planet onto the disk grow, that in turn
deepens the gap and lowers the accretion rate which is displayed in
Fig.~\ref{accq1.caption}.
Together these mechanisms effectively terminate the mass growth somewhere
between 5 and 10 $M_{Jup}$ which is entirely consistent with the observations
of extrasolar planets.
\subsection{Orbital and mass evolution of one planet}
\begin{figure}[ht]
\begin{center}
\epsfig{file=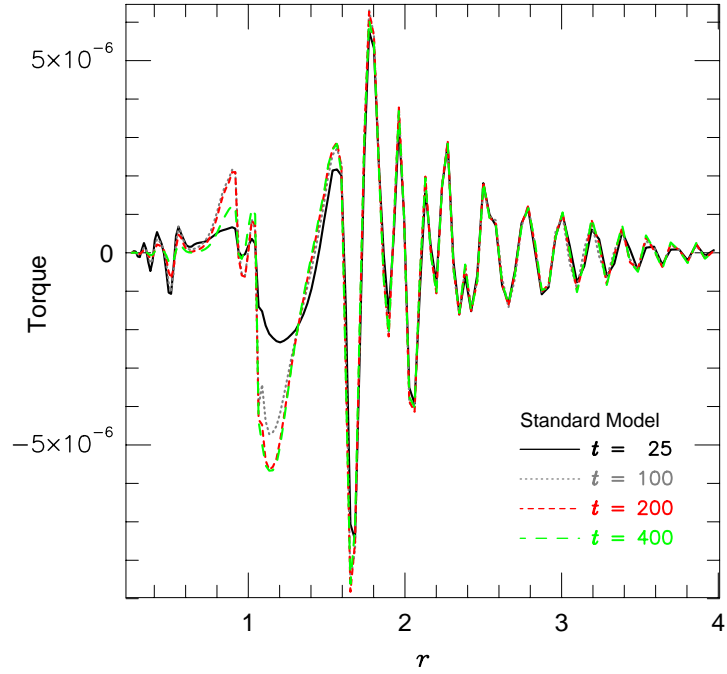, width=10cm}
\caption{\label{torque.caption} 
The radial dependence of the azimuthally averaged torque (in dimensionless
units) acting on the planet.}
\end{center}
\end{figure}

The orbital parameter of the planet in the previous calculations
were left unchanged even though the disk exerts gravitational torques
onto the planet which tend to change those. For a single planet in the disk
the change in the semi-major axis, $a$, may be computed to be
\[ 
 \dot{a} =  \frac{2}{a \, M_{pl} \, \Omega} \, \, T_\perp
\]
where $T_\perp$ is the $z$-component of the torque 
${\bf T} = {\bf r} \times {\bf F}$ acting on the planet
integrated over the disk's surface.
The radial distribution of the torque for the standard model is 
given in Fig.~\ref{torque.caption}.
As the direction of the motion is determined by the sign of the
total torque it is clear already from Fig.~\ref{torque.caption}
that the planet will migrate
inwards. The main contribution to this lowering of $a$ comes from
the density in trailing spiral arm near the outer edge of the gap. 
The deep minimum at $r\approx 1.62$ refers to the $2:1$ outer 
Lindblad resonance.
As the radial inner boundary of the computational domain was open to
mass outflow (supposed to have been accreted by the star), the contributions
of the regions inside the planets orbit are very small.

\begin{figure}[ht]
\begin{center}
\epsfig{file=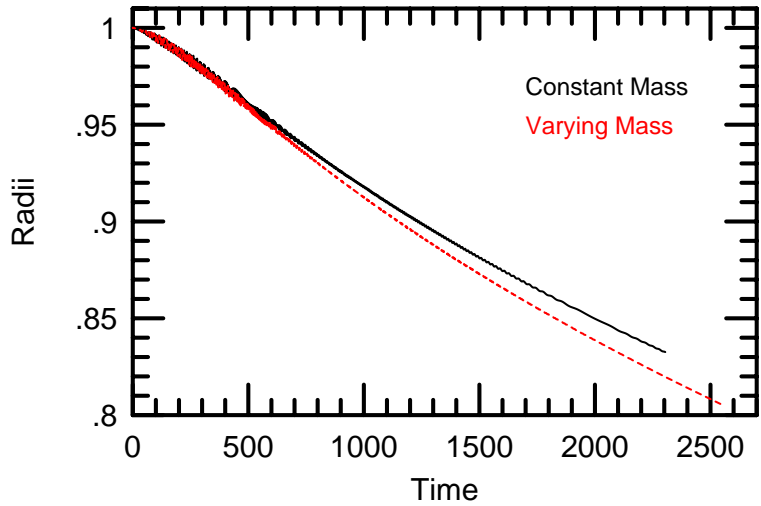, width=11cm}
\caption{\label{r1.caption}
The radial evolution of a migrating planet. In the first model (solid line)
the mass of the planet remained fixed while in the second case 
(dashed) the accreted mass was added to the planet.
}
\end{center}
\end{figure}

The change in the semi-major axis $\dot{a}$ can be translated into a
typical migration timescale
\[
        \tau_{mig} = \frac{a} {\left| \dot{a} \right|}
\]
and for the standard model we find that
$\tau_{mig} = 10^5 yr$ quite independent of numerical issues.
Hence, we do expect a rapid inward migration of the planet,
which is displayed in Fig.~\ref{r1.caption}.
The inclusion of the variation in the mass of the planet,
which has grown to about
$1.8 M_{Jup}$ within the 2500 orbits for the second case (dashed line),
changes the orbital evolution only slightly.
The eccentricity of the planets orbit does not
grow during the migration process. 
However, the migration time scale becomes longer during the
evolution and grows from $10^{5}yr$ to
nearly $2 \cdot 10^{5}yr$ at the end of the computation.
By this process of simultaneous inward migration and mass growth one can easily
explain the presence of massive planets closely to the central stars which are
on nearly circular orbits. 
In order that the planets do not migrate all the way inwards and are swallowed
by the stars a breaking mechanism is required \cite{lin1996}.
This may consist of a tidal
interaction of the planet with the star, or one may just 
assume that at some point the mass
of the accretion disk is exhausted which terminates the radial evolution
in a quite natural way.
\begin{figure}[ht]
\begin{center}
\epsfig{file=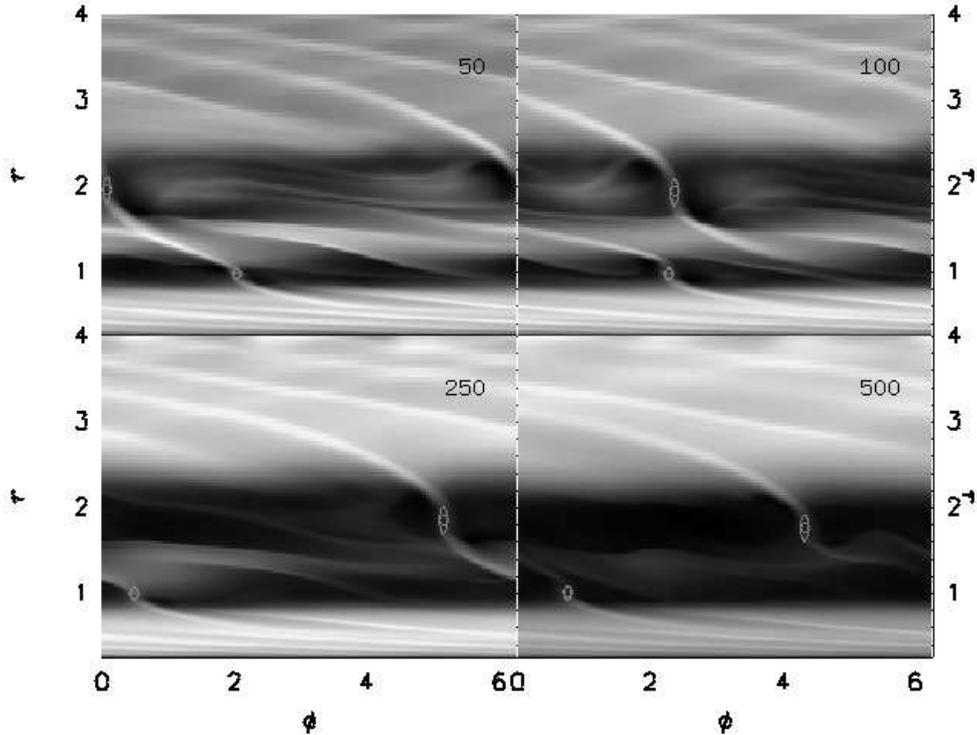, width=13cm}
\caption{\label{multi4.caption}
Gray scale plot of the surface density for the two-planet run at four
different times, given
in units of the initial orbital period of the inner planet.
The location of the planets and the sizes of their Roche lobes are indicated.
}
\end{center}
\end{figure}
\subsection{Evolution of two planets}
As it is believed that more eccentric orbits are caused by the interaction
of several objects, we have performed a run with two embedded Jupiter-type
planets. Initially they were located at $1 a_J$ and $2 a_J$ in opposition
to each other, i.e. separated by $\Delta \varphi = 180^{\circ}$.
The other physical parameter are identical to the standard model.
The system of the two planets and the star is treated as a three body system,
where the gravitational
back-reaction of the disk on the three objects is fully
taken into account. The origin of the coordinate system is
fixed in the central star and the equations of motion are
integrated using a fourth order Runge-Kutta method. 
The two planets were assumed to be on circular orbits initially.

Two planets create a much more complicated pattern of wave-like disturbances
in the density distribution of the disk than just one planet does.
This is displayed in Fig.~\ref{multi4.caption}.
In a calculation of one planet on a fixed circular orbit,
the wave pattern induced in the disk is stationary
in the co-rotating frame. As seen in the plot, here it changes strongly
with time. Clearly visible is also the clearing process of mass located
initially between the two planets, at about $t=500$ one common huge gap
has formed. At the same time the inner boundary at $r=.25 a_J$ is open
to mass outflow to the star, and the density is reduced in that region.

\begin{figure}[ht]
\begin{center}
\epsfig{file=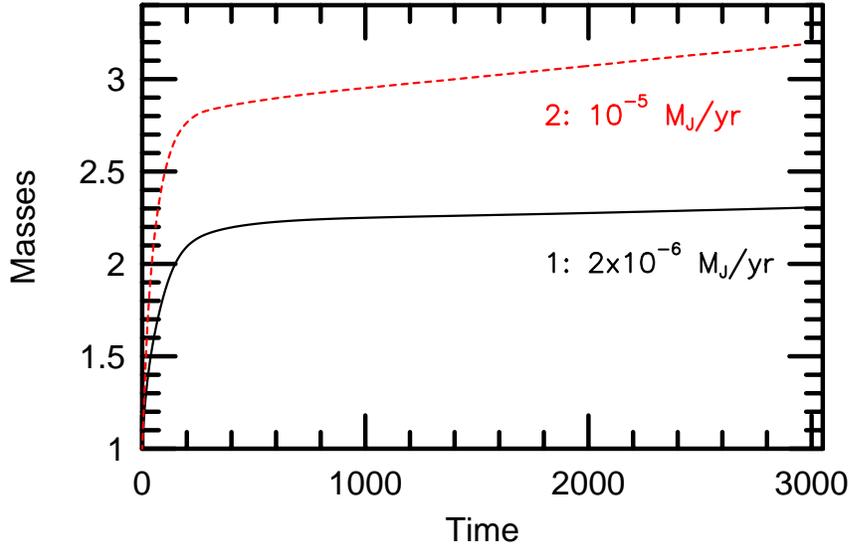, width=12cm}
\caption{\label{m2.caption}
Evolution of the masses of the two planets (1: inner planet, 2: outer planet).
The inferred mass accretion rates
during the longterm evolution are indicated.
Time is given in units of the initial period of the inner planet.
}
\end{center}
\end{figure}

As both planets accrete basically the mass which enters their Roche-lobes
their masses are growing in time but for the inner planet only that gas is
available which flows through the gap created by the outer one.
Hence the mass of the inner planet grows at a smaller rate than the outer
one, see Fig.~\ref{m2.caption}.
During the initial gap clearing process the planets grow
rapidly in mass and finally reach the asymptotic rates indicated.

\begin{figure}[ht]
\begin{center}
\epsfig{file=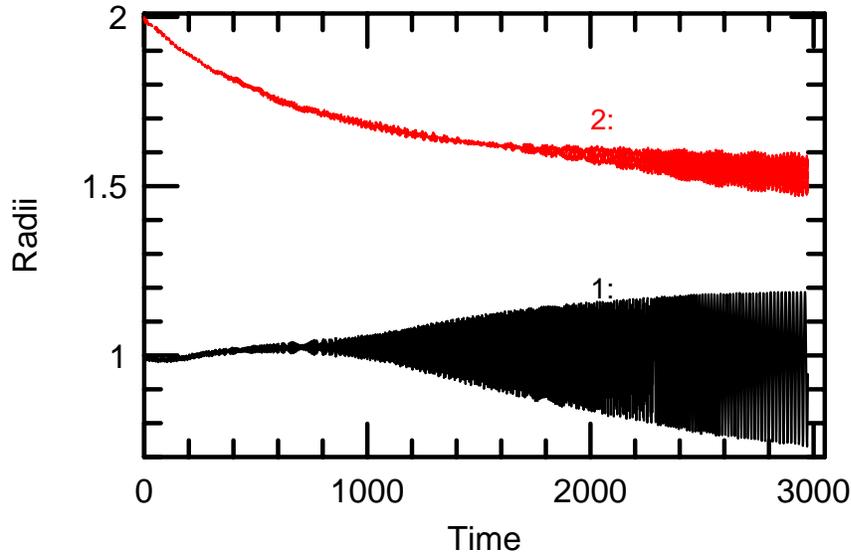, width=12cm}
\caption{\label{r2.caption}
Evolution of the semi-major axis of the two planets.
During the evolution the fluctuations of the radial distances to the
star strongly increases indicating a growth in eccentricity of the planet.
}
\end{center}
\end{figure}
At the same time the gravitational interaction of the two planets, the star 
and the disk lead to changes in the orbital elements of the objects.
The change in the semi-major axis of the planets is given in
Fig.~\ref{r2.caption}.
The outer planet behaves as in the previous case and migrates slowly inwards.
Visible is however, an increase in the eccentricity of the orbit which
has grown to about 0.05 at the end of the computation at $t=3000$.
The inner planet behaves completely different though: after a very
brief initial inward motion it moves outwards and remains there while 
its eccentricity $e_2$ grows strongly due to the interaction with
the approaching outer planet, reaching $e_2 = 0.2$  at $t=3000$.
The two planets eventually come so close to each other that
their orbits are no longer
stable and one of them may for example be ejected by the system leaving
just one planet on a highly eccentric orbit.
\section{Conclusion}
We have presented calculations of massive, Jupiter-type
planets embedded in protoplanetary disks.

By considering initially one planet on a fixed circular orbit, it was shown
that even though the planet opens up an annular gap in disk it is
nevertheless able to accrete more mass from its surroundings.
As more massive planets induce a wider and deeper gap the mass
accumulation basically terminates,
which puts the upper limit to the mass of the planet mass at about
5 to 10 $M_{Jup}$, in good agreement with the observations.

The net gravitational torque of the disk exerted
on the planet is generally negative
and induces an inward migration of the planet on timescales of about 
$t_{mig} \approx 10^{5} yr$. During this migration the planet remains
essentially on a circular orbit. Using this migration mechanism some basic
characteristics (Fig.~\ref{e-r.caption})
of the short period extrasolar planets may be explained.

By considering the evolution of a multi-planet system consisting of
two Jupiter-type planets we could show that by mutual gravitational
interaction the inward motion of the inner planet may come to a halt.
As the outer planet approaches the gravitational interaction between the two
planets increases which may render their orbits unstable, leading to systems
similar to $\upsilon$~And.
If the protoplanetary nebula has dissipated before
the planets come very close to each other one is left with a system of
massive planets at several AU distance, similar to our own Solar System.

Through numerical computations which model the interaction
between planets and the protoplanetary disk it has been shown that some basic
features of the physical parameter (masses, orbital elements) of the observed
extrasolar planets may be explainable.
To obtain additional insight, more elaborate three-dimensional
models which may include radiative effects are required. 

\begin{acknowledgment}
Computational resources of the Max-Planck Institute for Astronomy
in Heidelberg were available for this project and are gratefully
acknowledged.  This work was supported by the Max-Planck-Gesellschaft,
Grant No. 02160-361-TG74.
\end{acknowledgment}

\end{article}

\end{document}